\documentclass[showpacs,jcp,12pt]{revtex4-1}
\usepackage[english]{babel}
\usepackage{amsmath,graphicx,amssymb} 
\usepackage{times}
\usepackage{epsfig} 
\usepackage{epstopdf} 
\usepackage{color}

\begin{document} 
\title{Isotopic fractionation in proteins as a measure of hydrogen bond length
}

\author{Ross H. McKenzie}
\email{email: r.mckenzie@uq.edu.au}
\homepage{URL: condensedconcepts.blogspot.com}
\affiliation{School of Mathematics and Physics, University of Queensland,
  Brisbane 4072, Australia} 

\author{Bijyalaxmi Athokpam and Sai G. Ramesh}
\affiliation{Department of Inorganic and Physical Chemistry,
Indian Institute of Science, Bangalore 560 012, India}

\date{\today}
                   
\begin{abstract}
If a deuterated molecule containing strong intramolecular hydrogen bonds is
placed in a hydrogenated solvent it may preferentially exchange deuterium for
hydrogen. This preference is due to the difference between the vibrational
zero-point energy for hydrogen and deuterium.  It is found that the associated
fractionation factor $\Phi$  is correlated with the strength of the
intramolecular hydrogen bonds. This correlation has been used to determine the
length of the H-bonds (donor-acceptor separation) in a diverse range of enzymes
and has been argued to support the existence of short low-barrier H-bonds.
Starting with a potential energy surface based on a simple diabatic state model
for H-bonds we calculate $\Phi$ as a function of the proton donor-acceptor
distance $R$.  For numerical results, we use a parameterization 
of the model for symmetric
O-H$\cdots$O bonds [\emph{Chem. Phys. Lett.} {\bf 535}, 196
(2012)].  We consider the relative contributions of the O-H stretch vibration,
O-H bend vibrations (both in plane and out of plane), tunneling splitting
effects at finite temperature, and the secondary geometric isotope effect. We
compare our total $\Phi$ as a function of $R$ with NMR experimental results for
enzymes, and in particular with an empirical parametrization
$\Phi(R)$, used previously to determine bond lengths.
\end{abstract}

\pacs{}
\maketitle 

\section{Introduction}

The issue of low-barrier hydrogen bonds in proteins and whether they play any
functional role, particularly in enzyme catalysis, is controversial
\cite{Cleland94,Schutz04,Das06,Cleland10,Perrin10,Hosur13,Nadal14,Graham2014,Klinman2015,Nichols2015}.  Identifying
such short hydrogen bonds, characterised by a donor-acceptor distance of $R
\simeq 2.45-2.65$ \AA, is not completely straightforward \cite{Harris00}. In
protein X-ray crystallography, the standard errors in inter-atomic distances are
about 10 to 30 per cent of the resolution. Hence, for an X-ray structure with
$2.0$ \AA resolution, the standard errors in the distances are $\pm (0.2  -
0.6)$ \AA. This uncertainty makes it difficult to distinguish between short
strong bonds and the more common weak long bonds, with $ R > 2.8$ \AA \
\footnote{One might also consider whether the fact that X-ray crystal structures
	are refined with classical molecular dynamics using force fields that are
	parametrised for weak H-bonds may also be a problem. Such refinements may
naturally bias towards weak bonds, i.e., the longer bond lengths that are common
in proteins.}.  NMR provides an alternative method of bond length determination
via the $^1$H chemical shift. An independent  NMR ``ruler'' involves isotopic
fractionation, where one measures how much the relevant protons (H) exchange
with deuterium (D) in a solvent.
\begin{equation}
{\rm 
Pr-H + D}_{\rm solvent}
\rightleftharpoons
{\rm 
Pr-D + H}_{\rm solvent} 
\label{equilib}
\end{equation}
Here Pr-H denotes a protein with a proton in the relevant hydrogen bond.
The fractionation ratio can also be determined from UV spectroscopy \cite{Kreevoy80}.

The  fractionation ratio is equilibrium constant of Eq.~\eqref{equilib}:
\begin{equation}
\Phi \equiv \frac{[{\rm Pr-D}][{\rm H}_{\rm solvent}]}{ [{\rm Pr-H}][{\rm D}_{\rm solvent}]}
\label{phi}
\end{equation}
Translated into partition functions, $\Phi$ is essentially determined by the
relative zero-point energy (ZPE) of a D relative to an H in the protein. As
described by Kreevoy and Liang \cite{Kreevoy80}, the ratio is given by 
\begin{equation}
k_B T \ln \Phi = Z_{\rm H-Pr} - Z_{\rm D-Pr} +
Z_{\rm D,solvent} - Z_{\rm H,solvent}
\label{phi2}
\end{equation}
where $T$ is the temperature and $Z_{\rm H-Pr} $ denotes the 
zero-point energy of a proton participating in the relevant
hydrogen bond in the protein.
Throughout this paper we set $T=300$ K.

Fractionation is a purely quantum effect.
If the nuclear dynamics were classical, the fractionation ratio would be one.
It would also be one if there were no changes in the vibrational frequencies ---
more correctly, zero-point energies --- of both H and D when they moved from the
solvent to the protein. However, the vibrational potentials \emph{are} different
in the two environments.  The donor-acceptor distance, $R$, is typically shorter
in the protein, indicating a stronger H-bond and a softer X-H stretch potential
(X is the H-bond donor). Consequently, the difference between H and D
zero-point energies gets larger \cite{McKenzieJCP14}, and $\Phi$ gets smaller with
decreasing $R$. However, for very short bonds, typically when the donor and
acceptor share the H or D atoms, the stretch frequencies begin to harden and
$\Phi$ starts to increase. $\Phi$  then has a non-monotonic dependence
on $R$ \cite{Hibbert90}. 

Mildvan and collaborators \cite{Harris00,Mildvan02} considered a particular
parametrisation of the H-bond potential to connect the observed fractionation
ratio with donor-acceptor bond lengths in a range of proteins. They generally
find reasonable agreement  between determinations of the length from the fractionation
factor and that from the NMR
chemical shift. In particular, the uncertainty is less than that deduced from
X-rays.

In this paper, we systematically investigate how the fractionation factor $\Phi$
varies with the donor-acceptor distance $R$. Specifically, we consider the
relative importance of different contributions to $\Phi$.  We find that the
competing quantum effects associated with the X-H stretch and bend modes are the
most significant. Non-degeneracy of the two bend modes and tunnel splitting of
the stretch mode have small but noticeable effects.  The only important effect
of the secondary geometric isotope effect is that it enhances the contribution
from the tunnel splitting, mostly for $R \sim 2.4-2.6$ \AA. For most values of
$R$, the value of $\Phi$ we calculate differs from the empirical relation due to
Mildvan et al.~\cite{Mildvan02} that has been used to determine bond lengths in
enzymes.

\section{Method}
\label{sec:method}

We calculate the H/D zero-point energies, and hence $\Phi$, with the
electronic ground state potential of a two-diabatic state model
\cite{McKenzieCPL}. For X and Y as donor and acceptor, the two diabatic states
of the model are X-H$\cdots$Y and X$\cdots$H-Y, which are modelled as Morse
oscillators. The coupling between the diabats is a function of $R$, the X-Y distance,
 as well
as the H-X-Y and H-Y-X angles; it decreases exponentially with increasing R
 and gets weaker with larger angular excursions of the H atom.
Previously, we showed that this model can give a quantitative
description of the correlations observed \cite{Gilli} for a diverse range of
chemical compounds between $R$ and X-H bond lengths, vibrational frequencies,
and isotope effects \cite{McKenzieJCP14}.

We now briefly discuss the domain of applicability
of this simple model to hydrogen bonds in proteins, which are certainly
complex and diverse chemical systems.
First, our focus is on a small (but potentially important) sub-class of hydrogen bonds: short strong bonds. Second, we consider the simplest possible model that might capture the essential features of these bonds, independent of the
finer structural details of a specific protein.
The goal is to obtain physical insight into the different quantum
effects that contribute to the fractionation factor,
as well as their (non-monotonic) trends with donor-acceptor
distance.
H-bonds in proteins vary 
from weak to strong,  and can further be 
modified by coupling to other neighbouring H-bonds
\cite{Wang2014PNAS}. 
Also important are the proximity to and accessibility to the solvent,
and anisotropic electric fields arising from neighbouring charged amino acid
residues.  An example of the latter occurs in the Photoactive Yellow Protein
(PYP) where the existence of a possible low barrier H-bond may be dependent on
deprotonation of the neighbouring Arg52 residue \cite{Nadal14,Kanematsu2014}.  A
key feature of the two-diabatic state model used here is that it
takes the donor-acceptor bond distance and the pK$_a$ difference as the key bond
descriptors. These are input from available experimental information. These two
parameters will certainly be modified by chemical substituents \cite{Oltrogge},
solvent, and perturbations from the local electric field as indicated above.
Description of multiple H-bonds requires generalisation of the model
considered here to include more than two diabatic states \cite{McKenzieJCP}.

The parametrization used  in References \onlinecite{McKenzieJCP14}
and \onlinecite{McKenzieCPL} 
was for O-H$\cdots$O symmetric hydrogen bonds, i.e. the donor and acceptor have
the same proton affinity (pK$_a$). In the present work, we retain this
parametrization.  This is an approximation for comparisons with H-bonds in
proteins, which are generally asymmetric (donor and acceptor with different
pK$_a$).  Many H-bonds in proteins are actually
N-H$\cdots$O bonds. However, as the H-bonds become stronger 
($R \lesssim 2.5 $ \AA) 
the equal proton affinity 
approximation becomes more reliable. At such distances, the donor and acceptor
effectively share the H atom. 
In the diabatic state model, the off-diagonal
coupling element
becomes large enough to strongly suppress or eliminate the barrier for the H
atom transfer.  Kreevoy and Liang \cite{Kreevoy80}, Bao et al.\cite{Bao99},
and Oltrogge and Boxer\cite{Oltrogge}
considered how  asymmetry in the one-dimensional proton transfer potential modifies the
fractionation factor. Non-degeneracy smaller than 800 cm$^{-1}$ (or
equivalently, a pK$_a$ difference of about 2) has only a small effect on the proton
transfer potential and the fractionation factor when $R < 2.5 $ \AA.

The total vibrational zero-point energy for Pr-H/D is
\begin{equation} 
Z(R) \equiv
Z_\parallel(R) + Z_{\perp,o}(R)+ Z_{\perp,i}(R).
\label{eqn-zpe} 
\end{equation}
The three terms are the zero-point energy associated with X-H vibrations
parallel to the hydrogen bond (stretch), out-of-plane bend ($o$), and in-plane
bend ($i$) of X-H$\cdots$Y, respectively. (The plane typically refers to that of
X-H.) The simple summation in the above equation points to our assumption that
these modes are uncoupled.

The O-H/D stretch zero-point energy is calculated numerically by using the
sinc-function Discrete Variable Representation (DVR) \cite{Colbert92} to solve
the one-dimensional Schrodinger equation as a function of $R$. This gives an
essentially exact treatment for the significant anharmonic and tunnelling
effects that occur for low-barrier bonds \cite{McKenzieJCP14}. 

We treat zero-point energies of bend modes as half their classical harmonic
frequencies as a function of $R$. 
To break the degeneracy of the two modes, we use the result from the
model itself that hardening of the two bend motions is similar:  
\begin{equation} 
\Omega_{\perp,o/i}(R)^2 
=
\omega_{\perp,o/i}^2 + 2 f(R)
\label{eqn-bend} 
\end{equation}
where $\omega_{\perp,o/i}$ is the frequency in the absence of an H-bond and the
function $f(R)$ is given in Eqn. (6) of Ref.~\onlinecite{McKenzieCPL}. At least in the
$R$ range of interest, $f(R)$ is a positive function that monotonically
decreases with increasing $R$ (compare Figure 1).  
In general $ \omega_{\perp,i} >
\omega_{\perp,o}$ and so $ \Omega_{\perp,i} > \Omega_{\perp,o}$. 
Here we take 
$\omega_{\perp,o}$ = 650 cm$^{-1}$ and
 $\omega_{\perp,i} = $ 650 or 1600 cm$^{-1}$. 
The $\omega$
parameters
for the deuterium isotope are taken to be $\sqrt{2}$ smaller than for
the H isotope.

Eq.~\eqref{phi2} for $\Phi$ assumes that only the ground state energies of the
species is relevant (at the temperature of interest). For a symmetric proton
transfer potential, one expects a first excited state due to tunnel splitting
that would be close in energy to the ground state. In our model, this
appears along the X-H stretch coordinate. Hence the first excited state for the
H/D motion makes a further multiplicative contribution of the form 
\begin{equation}
\Phi_{\rm tun} = \frac{1+\exp(-\delta E_D/k_BT)}{1+\exp(-\delta E_H/k_BT)}
\label{eqn-tun}
\end{equation}
to the fractionation
factor, where $\delta E_{H/D} \equiv E_{0^-} - E_{0^+}$ is the tunnel splitting.

Another contribution to $\Phi$ comes from the secondary geometric isotope effect (SGIE)
where the X-Y distance changes upon deuteration. This is a subtle effect with a
non-monotonic dependence on $R$ \cite{Ichikawa00,Li11,McKenzieJCP14}. As shown
in Ref.~\onlinecite{McKenzieJCP14} and references therein, it arises because the
rates of change with $R$ of the zero-point energy for H and D are different
(compare Figure 1).
The net effect is that true minima
of the total system energy with respect to $R$
 for both H and D ($R_H$ and $R_D$) are shifted relative
to the classical minimum ($R_o$). The difference between the minima is largely
under 0.04 $\AA$ with $R_D>R_H$, but the resulting effects on frequencies are
substantial. There are two consequences of SGIE for the fractionation ratio.
First, the zero point energy for the H and D should be calculated at their
respective minima. Second, an elastic energy associated with the stretching of
the donor-acceptor distance, of the form $\tfrac{1}{2}K(R_{H/D}-R_o)^2$, must be
included. Here, $K$ is the elastic constant that is
parametrised empirically in Ref. \onlinecite{McKenzieJCP14}.
We have included both consequences of the SGIE in our calculation.

The above details describe the calculation of the [Pr-D]/[Pr-H] part of $\Phi$,
in Eq.~(\ref{phi}), as a
function of $R$. The corresponding ratio for the solvent 
is taken to be the calculated model value at $R = 2.8$
\AA, approximately the relevant length in water.
Later we discuss how our results are not particularly sensitive to this
exact choice of a reference distance.

\section{Results and Discussion}

\begin{figure}[htb] 
\centering 
\includegraphics[width=84mm]{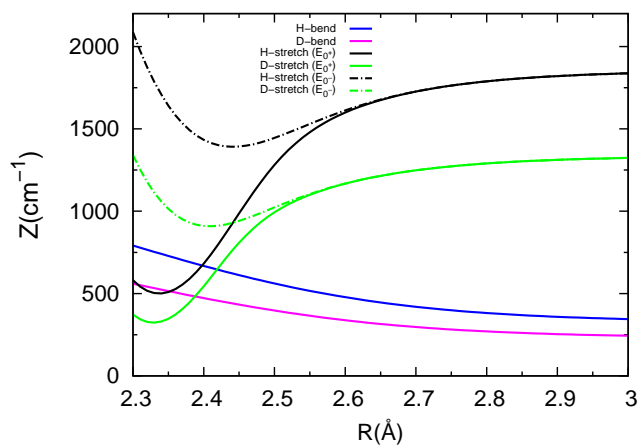}
\caption{Competing quantum effects.
The zero-point energies of the out-of-plane bend mode and of the two lowest stretch quantum states (due to tunnel splitting, $E_{0^\pm}$)
are shown for both H and D isotopes  as a function of the donor-acceptor distance $R$. The black curves [H-stretch ($E_{0^\pm}$)] correspond to the stretch mode of 
the H isotope, while green curves [D-stretch ($E_{0^\pm}$)] are those of the D isotope (solid: $+$, dot-dashed: $-$). Blue (H-bend) and magenta (D-bend) curves are 
out of plane bend zero-point energies for the H and D isotopes, respectively.
With increasing $R$, the stretch zero-point energies increase and those of the bend decrease. Note how for $R < 2.55 $ \AA \  (2.45 \AA), 
the tunnel splitting of the stretch mode for the H (D) isotope becomes observable.
}
\label{fig1}
\end{figure}

\begin{figure}[htb] 
\centering 
\includegraphics[width=84mm]{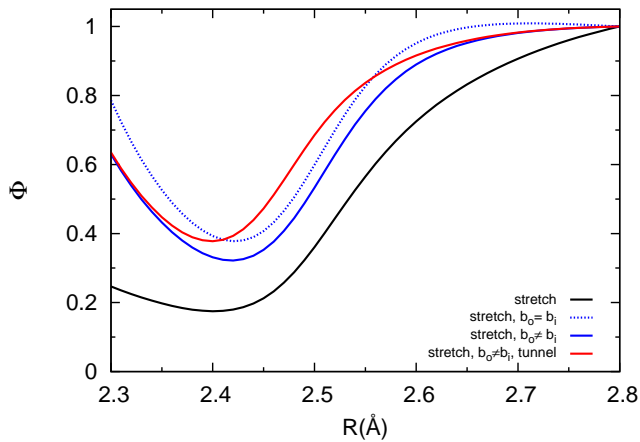}
\caption{
Effect of the bend modes and tunnel splitting on the fractionation ratio. The black curve (stretch) includes solely the effects of the X-H stretch vibrational mode. The blue curves include the effect of the in-plane (b$_i$) and out-of-plane (b$_o$) bending vibrational modes. The upper dashed curve (stretch, b$_{o}=$ b$_{i}$) is for degenerate bend modes, while the lower solid curve (stretch, b$_{o}\neq $ b$_{i}$, tunnel) includes the contribution from the first excited X-H stretch state (tunnel splitting) for non-degenerate bend modes.
}
\label{fig2}
\end{figure}

\begin{figure}[htb] 
\centering 
\includegraphics[width=84mm]{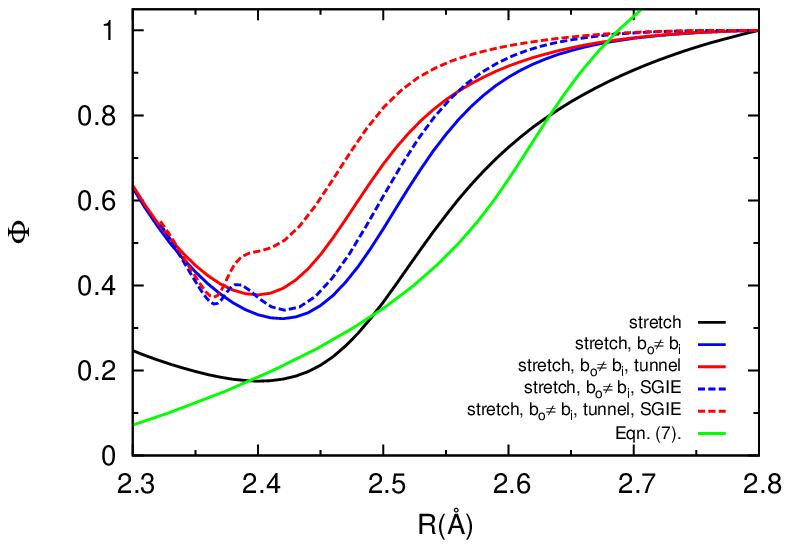}
\caption{
Corrections due to the secondary geometric isotope effect (SGIE) are shown as
the dashed curves: the blue dashed line (stretch, b$_{o}\neq$ b$_{i}$, SGIE) is without including tunneling and the red dashed line (stretch, b$_{o}\neq$ b$_{i}$, tunnel, SGIE) is with tunneling. 
The solid curves of the same colour are from Figure \ref{fig2}, which do not include the SGIE.  The green curve is the empirical function, defined by Eqn.~(\ref{mildvan}), and given in Ref. \onlinecite{Mildvan02}. Note that the slight undulation in the dashed curves below 2.4 \AA \ can be traced to the very rapid change of the SGIE ($R_D-R_H$) in that region; see the lower panel of Figure 7 in Ref.~\onlinecite{McKenzieJCP14}.
}
\label{fig3}
\end{figure}

\subsection{Role of competing quantum effects}

Figure \ref{fig1} shows the computed trends of the stretch and bend zero-point
energies from the model for both H and D isotopes.
The solid black (H) and green (D) curves show the O-H/D stretch zero-point
energies. The limiting energies at large $R$ are those of O-H and O-D bonds,
about 1800 and 1300 cm$^{-1}$, respectively. Of relevance to the fractionation
factor $\Phi$ is the \emph{difference} between these curves, which decreases
(for the most part) with decreasing $R$. From Eq.~\eqref{phi2}, this leads to a
drop in $\Phi$. The black curve in Figure \ref{fig2} shows the fractionation
ratio with only the O-H/D stretch zero-points included.

A countervailing influence on $\Phi(R)$ comes 
from the bends, which harden in frequency with
decreasing $R$. Figure \ref{fig1} shows the zero-points only for the out-of-plane
bends for the H and D cases. Their limiting values at large $R$ are
$\tfrac{1}{2}\omega_{H,\perp,o}$ (=325 cm$^{-1}$) and
$\tfrac{1}{2\sqrt{2}}\omega_{H,\perp,o}$ (=230 cm$^{-1}$),
respectively, where $\omega_{H,\perp,o}=650$ cm$^{-1}$. The corresponding trends
for the in-plane bend (not plotted) are obtained from Eq.~\eqref{eqn-bend},
setting $\omega_{H,\perp,i}=1600$ cm$^{-1}$. The consequences of the opposite
trends for the hydrogen and deuterium bend and stretch frequencies --- more
compactly, competing quantum effects --- has been the subject of much recent
study \cite{Habershon09,Li11,Markland12,Romanelli13,McKenzieJCP14,WangJCP14}.

Presently, for $\Phi$, it is $Z_H-Z_D$ that matters, which evidently also
showcases the competition between the X-H bends and the X-H stretch. The solid
blue curve in Figure \ref{fig2} shows that the hardening of the bend modes with
decreasing $R$ significantly increases the fractionation ratio compared to the
contribution from just the stretch mode.  This is one of the main results of
this paper.  Kreevoy and Liang \cite{Kreevoy80} previously pointed out that
bending modes could alter their results for the correlation of $\Phi$ and $R$.
They gave the rough estimate that $\Phi$ could be increased by a factor of about
1.7.  Edison, Weinhold, and Markley also mentioned the effect of the bend
modes \cite{Edison95}, finding values of $\Phi > 1$ for weak bonds.

\subsection{Role of non-degeneracy of the bend modes}

For the main results of this paper, the bend modes are made non-degenerate
using Eq.~\eqref{eqn-bend} since the out-of-plane and in-plane modes are
different in frequency in reality. We briefly investigate how different the
fractionation factor would be if the degeneracy were retained.

The dotted blue curve in Figure \ref{fig2} shows the plot of $\Phi$ with both
bends having frequencies corresponding to the out-of-plane mode.  Evidently, the
change is relatively small, but not negligible. As per the model, the smaller
the bend frequency, the faster it hardens; $\Omega_{\perp}(R)/\omega_{\perp} =
\sqrt{1+ 2f(R)/\omega^2_{\perp}}$. Hence, changing one of the bends frequencies to
1600 cm$^{-1}$ reduces $Z_H-Z_D$ relative to the degenerate case. Bend
non-degeneracy can reduce the fractionation by about 10-20\% compared to the
degenerate case.

\subsection{Role of the tunnel splitting}

For long H-bonds, the proton transfer potential has a high barrier and the
tunnel splitting of the vibrational ground state is negligible
\cite{McKenzieJCP14}. However, for $R < 2.55 \ \AA$, the splitting becomes
significant, as can be seen in Figure \ref{fig1}. A multiplicative correction
$\Phi_{tun}$ (Eq.~\eqref{eqn-tun}) introduces the effect of the thermal
population of the first low-lying excited state. This factor is always larger
than one because the H tunneling splitting is larger; $\delta E_{\rm H} > \delta
E_{\rm D}$. When the tunnel splitting is much larger than the thermal energy
$k_B T$ (i.e. for $R < 2.4$ \AA) the correction factor is extremely close to
unity. When the tunnel splitting is much less than $k_B T$ the correction
factor is approximately $(1 + (\delta E_{\rm H}-\delta E_{\rm D})/2k_B T)$.

Figure \ref{fig2} shows that the tunnel splitting has a small but non-negligible
effect in the range, $R \sim 2.4-2.6$ \ \AA.  We should also clarify the
nomenclature here. For sufficiently small $R$, the barrier is no longer present,
and so there is strictly no ``tunnel splitting." We just have two well-separated
vibrational energies instead.  Note that in a solvent there will be local
dynamical fluctuations of the local electric field that couple to the electric
dipole moment associated with the X-H stretch and for large enough fluctuations
the tunnel splitting will not appear because it will be destroyed by quantum
decoherence \cite{Bothma10}. Also, when the proton affinity of the donor and
acceptor differ by more than about 500 cm$^{-1}$ (1.5 kcal/mol or 1.3 pK$_a$ units) this effect may
be absent.

\subsection{Role of the secondary geometric isotope effect}

The SGIE has a significant effect on the stretch mode vibrational frequencies
for $R \sim 2.4-2.5$ \AA, where the proton transfer barrier has effectively
disappeared. Its inclusion yielded better agreement of the H/D stretch frequency
ratio; compare Figure 8 in Ref.~\onlinecite{McKenzieJCP14}, where this ratio is
1 (1.15) with (without) SGIE, which is a sizeable change for strong short
H-bonds.

Figure \ref{fig3}, however, points to a more modest influence of the
SGIE on the
fractionation ratio. The only significant effect is how it modifies the
contribution from the tunnel splitting.  Without the SGIE,
the correction factor is approximately $(1 + (\delta E_H(R)-\delta E_D(R))/2k_B
T)$.  With the SGIE, the correction factor is approximately $(1 + (\delta
E_H(R_H)-\delta E_D(R_D))/2k_B T)$. This is larger because $R_D > R_H$, and an
increase in donor-acceptor distance of as little as $0.02$ {\AA} for D relative
to H can increase the energy barrier and thereby noticeably decrease $\delta
E_D(R_D)$ \cite{McKenzieJCP14}.

If tunneling contributions are suppressed by, e.g., a sizeable
difference in $pK_a$'s between the donor and acceptor, the above analysis
suggests that the SGIE would have only a small influence on the fractionation
factor. The key effect that appears to govern the magnitude range of $\Phi(R)$
according to our model is the competing quantum effect between the X-H stretch
and X-H bends. 

\subsection{Sensitivity to choice of reference distance}

Calculation  of $\Phi$,
in Eqn. (\ref{phi}) requires knowledge of the ratio of H/D concentration
in the solvent. In the reported calculations
we took this ratio to be given  by the value calculated within our model 
at $R = 2.8$
\AA, approximately the relevant length in water.
Our results are not particularly sensitive to this
exact choice of this reference distance.
For $R > 2.7 $ \ \AA \ the difference between the H and D zero-point
energies is small. This can be seen from Figure \ref{fig2}.
Even for the case of purely stretch modes taking the reference
distance to be $R=2.7 $ \
\AA \ would only increase $\Phi(R)$ by
about 10 per cent compared to the values shown in our curves.

\subsection{Comparison with experiment}

To put our results in context we now briefly review 
previous measurements of $\Phi$ that have been used to 
deduce a value for $R$ in a specific molecule.

Based on calculations from an empirical one-dimensional quartic potential \cite{Bao99},
Mildvan et al.\cite{Mildvan02} considered a relation between the fractionation 
factor and donor-acceptor distance,
\begin{equation}
R = (2.222 + 1.192 \Phi - 1.335 \Phi^2 + 0.608 \Phi^3) \ {\rm \AA}.
\label{mildvan}
\end{equation}
It was used together
with measurements of fractionation ratios for 18 H-bonds
in several different enzymes to deduce the length $R$.  The values they obtained
for $R$ were mostly in agreement with values of $R$ deduced from NMR chemical
shifts, and from X-ray crystallography.  Values of $\Phi$ ranged from 0.32 to
0.97 and the corresponding values of $R$ were in the range 2.49 to 2.68 \AA.
However, Figure \ref{fig3} shows  significant differences between equation
(\ref{mildvan}) and our results. 

Klug et al.\cite{Klug97} studied crystals of
the dihydrated sodium salt of hydrogen bis(4-nitrophenoxide)
and found a fractionation ratio
of $0.63 \pm 0.04$. Using the Kreevoy and Liang \cite{Kreevoy80}
 parametrisation, they noted that
this value was inconsistent with the bond length observed via X-rays,
$R=2.452$ \AA, and with the value of $\Phi = 0.31 \pm 0.03$ deduced from
UV spectroscopy for 
bis(4-nitrophenoxide) in acetonitrile solution.
Consequently, they suggested that ``the solid was not in isotopic equilibrium
with the solvent from which it was precipitated.''
However, their results are consistent with our parameterisation
of $\Phi$ versus $R$, if tunnel splitting is not included.

Loh and Markley\cite{Loh94}  found fractionation factors in the range
0.28-1.47 for the different H-bonds in the protein
 staphylococcal nuclease.
Clearly, we cannot explain their $\Phi$ values larger than unity.
However, it should be pointed out that many of these bonds are weak
with donor-acceptor distances in the range, $R \sim 2.8 - 3.3$ \ \AA \,
and that no clear correlation was observed between the values of $\Phi$ and $R$.
Loh and Markley did note a difference between the fractionation factors of backbone amide
bonds that are solvent-exposed (average value $0.98$) and those that are not
(average value $0.79$).
This observation is consistent with 
what one might anticipate: solvent accessibilility could affect the 
the effective pK$_a$ of the donor and/or
acceptor. A consequent change (increase) in distance between them may also ensue.
These would lead to a weakening of the H-bond, and therefore move $\Phi$ values
closer to 1. In the model we employ, this solvent effect
would enter parametrically as a difference in effective pK$_a$ and donor-acceptor distance.
However, the works of Khare \emph{et al.} \cite{Khare1999} and LiWang and Bax
\cite{Liwang1996} point to minimal effects. The former, which reported on the
immunoglobulin G binding domains of protein G, found little difference between average
fractionation factors (average $\Phi$
of 1.05 for $\alpha$-helical, 1.13 for $\beta$-sheet, and 1.08 for solvent-exposed
residues). LiWang and Bax gave similar findings for ubiquitin. 
A recent study of the core of protein Kinase A also
found no correlation between $\Phi$ values and secondary structure \cite{Li2015}.

Thakur et al.\cite{Thakur13} have recently developed a new method for the
rapid determination of H/D exchange from two-dimensional NMR spectra.
Section S5 of their Supplementary material shows fractionation values for
three different proteins.
For 80 different amino acid residues in Tim23, the values 
ranged from 0.81 to 1.73.
For 58 different amino acid residues in Ubiquitin, the values 
ranged from 0.34 to 1.67.
For 54 different amino acid residues in Dph4, the values 
ranged from 0.45 to 2.04.

Recently, an extensive study was made of mutants of the Green Fluorescent Protein
with a  short H-bond between the chromophore and the amino acid Asp148 \cite{Oltrogge}. 
The donor-acceptor bond length estimated from
X-ray structures was  $2.4 \pm 0.2 $ \ \AA. 
The pK$_a$ of the chromophore was systematically varied by 3.5 units
through halogen substitutions. This range covers the pK$_a$ matching
(degenerate diabatic states) required for strongest bonds \cite{McKenzieCPL}.
The experimental results were compared to calculations based
on a one-dimensional proton transfer potential based
on same diabatic state model used here.
The measured fractionation factors (deduced from analysis of UV absorption spectra)
 were in the range
 0.54 - 0.9, taking a minimum value for pK$_a$ matching.
This observation and a value of $\Phi=0.54$ for $ R=2.4 \pm 0.2 $ \ \AA \
are consistent with our analysis when the bend modes and tunnel
splitting are taken into account.

Edison, Weinhold, and Markley performed {\it ab initio} calculations 
for a wide range of peptide clusters \cite{Edison95}. They observed
a correlation between the fractionation ratio and the donor-acceptor 
distance.
For $R > 2.55$ \AA, the fractionation was larger than one, and
for $R \simeq 2.45$ \AA, $\Phi \simeq 0.6$.

\section{Conclusions}

We have shown that the H/D fractionation factor $\Phi$ is quite sensitive
to the donor-acceptor distance $R$ in hydrogen bonds, and so, in principle,
can be used as a ``ruler'' for determining bond lengths. 
However, caution is in order because there are a number of
subtle effects that modify the exact form of the relationship between
$\Phi$ and $R$. These include competing quantum effects between stretch and 
bend modes, non-degeneracy of the bend modes, tunnel splitting, the secondary
geometric isotope effect, and differences between the proton affinity of the donor
and acceptor.

Our results raise questions about whether values of $\Phi$ as small as 0.3 are
really possible for short bonds, contrary to some measurements and previous theoretical claims.
Equally, our results cannot explain $\Phi$ values that are much larger than 1 for long bonds.
The discrepancy for short bonds may be due to our assumption that the stretch and bend modes
are independent.  Although our model quantitatively describes many experimental
results for bond lengths, vibrational frequencies, and isotope effects, for $R
\sim 2.45 $ \ \AA, it does give stretch mode frequencies that are higher than
observed.  (See Figure 6 in  Ref. \onlinecite{McKenzieJCP14} and the associated
discussion.) This would also lead to a larger fractionation factor than observed.  Addressing
this issue will require a systematic investigation of solutions to the
vibrational Schrodinger equation for a higher-dimensional (probably four
dimensional) potential energy surface.  We leave that for a future study.

\begin{acknowledgments}

We thank Michael Ashfold, Steve Boxer, Stephen Fried, Tom Markland, Seth Olsen,
 Luke Oltrogge, and Lu Wang for helpful
discussions.  RHM received financial support from an Australian Research Council
Discovery Project grant.

\end{acknowledgments}

\bibliographystyle{apsrev4-1}
\bibliography{mckenzie4}

\begin{thebibliography}{38}%
\makeatletter
\providecommand \@ifxundefined [1]{%
 \@ifx{#1\undefined}
}%
\providecommand \@ifnum [1]{%
 \ifnum #1\expandafter \@firstoftwo
 \else \expandafter \@secondoftwo
 \fi
}%
\providecommand \@ifx [1]{%
 \ifx #1\expandafter \@firstoftwo
 \else \expandafter \@secondoftwo
 \fi
}%
\providecommand \natexlab [1]{#1}%
\providecommand \enquote  [1]{``#1''}%
\providecommand \bibnamefont  [1]{#1}%
\providecommand \bibfnamefont [1]{#1}%
\providecommand \citenamefont [1]{#1}%
\providecommand \href@noop [0]{\@secondoftwo}%
\providecommand \href [0]{\begingroup \@sanitize@url \@href}%
\providecommand \@href[1]{\@@startlink{#1}\@@href}%
\providecommand \@@href[1]{\endgroup#1\@@endlink}%
\providecommand \@sanitize@url [0]{\catcode `\\12\catcode `\$12\catcode
  `\&12\catcode `\#12\catcode `\^12\catcode `\_12\catcode `\%12\relax}%
\providecommand \@@startlink[1]{}%
\providecommand \@@endlink[0]{}%
\providecommand \url  [0]{\begingroup\@sanitize@url \@url }%
\providecommand \@url [1]{\endgroup\@href {#1}{\urlprefix }}%
\providecommand \urlprefix  [0]{URL }%
\providecommand \Eprint [0]{\href }%
\providecommand \doibase [0]{http://dx.doi.org/}%
\providecommand \selectlanguage [0]{\@gobble}%
\providecommand \bibinfo  [0]{\@secondoftwo}%
\providecommand \bibfield  [0]{\@secondoftwo}%
\providecommand \translation [1]{[#1]}%
\providecommand \BibitemOpen [0]{}%
\providecommand \bibitemStop [0]{}%
\providecommand \bibitemNoStop [0]{.\EOS\space}%
\providecommand \EOS [0]{\spacefactor3000\relax}%
\providecommand \BibitemShut  [1]{\csname bibitem#1\endcsname}%
\let\auto@bib@innerbib\@empty
\bibitem [{\citenamefont {Cleland}\ and\ \citenamefont
  {Kreevoy}(1994)}]{Cleland94}%
  \BibitemOpen
  \bibfield  {author} {\bibinfo {author} {\bibfnamefont {W.}~\bibnamefont
  {Cleland}}\ and\ \bibinfo {author} {\bibfnamefont {M.}~\bibnamefont
  {Kreevoy}},\ }\href@noop {} {\bibfield  {journal} {\bibinfo  {journal}
  {Science}\ }\textbf {\bibinfo {volume} {264}},\ \bibinfo {pages} {1887}
  (\bibinfo {year} {1994})}\BibitemShut {NoStop}%
\bibitem [{\citenamefont {Schutz}\ and\ \citenamefont
  {Warshel}(2004)}]{Schutz04}%
  \BibitemOpen
  \bibfield  {author} {\bibinfo {author} {\bibfnamefont {C.~N.}\ \bibnamefont
  {Schutz}}\ and\ \bibinfo {author} {\bibfnamefont {A.}~\bibnamefont
  {Warshel}},\ }\href@noop {} {\bibfield  {journal} {\bibinfo  {journal}
  {Proteins: Structure, Function, and Bioinformatics}\ }\textbf {\bibinfo
  {volume} {55}},\ \bibinfo {pages} {711} (\bibinfo {year} {2004})}\BibitemShut
  {NoStop}%
\bibitem [{\citenamefont {Das}\ \emph {et~al.}(2006)\citenamefont {Das},
  \citenamefont {Prashar}, \citenamefont {Mahale}, \citenamefont {Serre},
  \citenamefont {Ferrer},\ and\ \citenamefont {Hosur}}]{Das06}%
  \BibitemOpen
  \bibfield  {author} {\bibinfo {author} {\bibfnamefont {A.}~\bibnamefont
  {Das}}, \bibinfo {author} {\bibfnamefont {V.}~\bibnamefont {Prashar}},
  \bibinfo {author} {\bibfnamefont {S.}~\bibnamefont {Mahale}}, \bibinfo
  {author} {\bibfnamefont {L.}~\bibnamefont {Serre}}, \bibinfo {author}
  {\bibfnamefont {J.~L.}\ \bibnamefont {Ferrer}}, \ and\ \bibinfo {author}
  {\bibfnamefont {M.~V.}\ \bibnamefont {Hosur}},\ }\href@noop {} {\bibfield
  {journal} {\bibinfo  {journal} {Proceedings of the National Academy of
  Sciences}\ }\textbf {\bibinfo {volume} {103}},\ \bibinfo {pages} {18464}
  (\bibinfo {year} {2006})}\BibitemShut {NoStop}%
\bibitem [{\citenamefont {Cleland}(2010)}]{Cleland10}%
  \BibitemOpen
  \bibfield  {author} {\bibinfo {author} {\bibfnamefont {W.~W.}\ \bibnamefont
  {Cleland}},\ }\href@noop {} {\bibfield  {journal} {\bibinfo  {journal}
  {Advances in Physical Organic Chemistry}\ }\textbf {\bibinfo {volume} {44}},\
  \bibinfo {pages} {1} (\bibinfo {year} {2010})}\BibitemShut {NoStop}%
\bibitem [{\citenamefont {Perrin}(2010)}]{Perrin10}%
  \BibitemOpen
  \bibfield  {author} {\bibinfo {author} {\bibfnamefont {C.}~\bibnamefont
  {Perrin}},\ }\href@noop {} {\bibfield  {journal} {\bibinfo  {journal}
  {Accounts of Chemical Research}\ }\textbf {\bibinfo {volume} {43}},\ \bibinfo
  {pages} {1550} (\bibinfo {year} {2010})}\BibitemShut {NoStop}%
\bibitem [{\citenamefont {Hosur}\ \emph {et~al.}(2013)\citenamefont {Hosur},
  \citenamefont {Chitra}, \citenamefont {Hegde}, \citenamefont {Choudhury},
  \citenamefont {Das},\ and\ \citenamefont {Hosur}}]{Hosur13}%
  \BibitemOpen
  \bibfield  {author} {\bibinfo {author} {\bibfnamefont {M.}~\bibnamefont
  {Hosur}}, \bibinfo {author} {\bibfnamefont {R.}~\bibnamefont {Chitra}},
  \bibinfo {author} {\bibfnamefont {S.}~\bibnamefont {Hegde}}, \bibinfo
  {author} {\bibfnamefont {R.}~\bibnamefont {Choudhury}}, \bibinfo {author}
  {\bibfnamefont {A.}~\bibnamefont {Das}}, \ and\ \bibinfo {author}
  {\bibfnamefont {R.}~\bibnamefont {Hosur}},\ }\href@noop {} {\bibfield
  {journal} {\bibinfo  {journal} {Crystallography Reviews}\ }\textbf {\bibinfo
  {volume} {19}},\ \bibinfo {pages} {3} (\bibinfo {year} {2013})}\BibitemShut
  {NoStop}%
\bibitem [{\citenamefont {Nadal-Ferret}\ \emph {et~al.}(2014)\citenamefont
  {Nadal-Ferret}, \citenamefont {Gelabert}, \citenamefont {Moreno},\ and\
  \citenamefont {Lluch}}]{Nadal14}%
  \BibitemOpen
  \bibfield  {author} {\bibinfo {author} {\bibfnamefont {M.}~\bibnamefont
  {Nadal-Ferret}}, \bibinfo {author} {\bibfnamefont {R.}~\bibnamefont
  {Gelabert}}, \bibinfo {author} {\bibfnamefont {M.}~\bibnamefont {Moreno}}, \
  and\ \bibinfo {author} {\bibfnamefont {J.~M.}\ \bibnamefont {Lluch}},\
  }\href@noop {} {\bibfield  {journal} {\bibinfo  {journal} {Journal of the
  American Chemical Society}\ }\textbf {\bibinfo {volume} {136}},\ \bibinfo
  {pages} {3542} (\bibinfo {year} {2014})}\BibitemShut {NoStop}%
\bibitem [{\citenamefont {Graham}\ \emph {et~al.}(2014)\citenamefont {Graham},
  \citenamefont {Buytendyk}, \citenamefont {Wang}, \citenamefont {Bowen},\ and\
  \citenamefont {Collins}}]{Graham2014}%
  \BibitemOpen
  \bibfield  {author} {\bibinfo {author} {\bibfnamefont {J.~D.}\ \bibnamefont
  {Graham}}, \bibinfo {author} {\bibfnamefont {A.~M.}\ \bibnamefont
  {Buytendyk}}, \bibinfo {author} {\bibfnamefont {D.}~\bibnamefont {Wang}},
  \bibinfo {author} {\bibfnamefont {K.~H.}\ \bibnamefont {Bowen}}, \ and\
  \bibinfo {author} {\bibfnamefont {K.~D.}\ \bibnamefont {Collins}},\
  }\href@noop {} {\bibfield  {journal} {\bibinfo  {journal} {Biochemistry}\
  }\textbf {\bibinfo {volume} {53}},\ \bibinfo {pages} {344} (\bibinfo {year}
  {2014})}\BibitemShut {NoStop}%
\bibitem [{\citenamefont {Klinman}(2015)}]{Klinman2015}%
  \BibitemOpen
  \bibfield  {author} {\bibinfo {author} {\bibfnamefont {J.~P.}\ \bibnamefont
  {Klinman}},\ }\href@noop {} {\bibfield  {journal} {\bibinfo  {journal} {ACS
  Central Science}\ }\textbf {\bibinfo {volume} {1}},\ \bibinfo {pages} {115}
  (\bibinfo {year} {2015})}\BibitemShut {NoStop}%
\bibitem [{\citenamefont {Nichols}\ \emph {et~al.}(2015)\citenamefont
  {Nichols}, \citenamefont {Hargis}, \citenamefont {Sanishvili}, \citenamefont
  {Jaishankar}, \citenamefont {Defrees}, \citenamefont {Smith}, \citenamefont
  {Wang}, \citenamefont {Prati}, \citenamefont {Renslo}, \citenamefont
  {Woodcock},\ and\ \citenamefont {Chen}}]{Nichols2015}%
  \BibitemOpen
  \bibfield  {author} {\bibinfo {author} {\bibfnamefont {D.~A.}\ \bibnamefont
  {Nichols}}, \bibinfo {author} {\bibfnamefont {J.~C.}\ \bibnamefont {Hargis}},
  \bibinfo {author} {\bibfnamefont {R.}~\bibnamefont {Sanishvili}}, \bibinfo
  {author} {\bibfnamefont {P.}~\bibnamefont {Jaishankar}}, \bibinfo {author}
  {\bibfnamefont {K.}~\bibnamefont {Defrees}}, \bibinfo {author} {\bibfnamefont
  {E.~W.}\ \bibnamefont {Smith}}, \bibinfo {author} {\bibfnamefont {K.~K.}\
  \bibnamefont {Wang}}, \bibinfo {author} {\bibfnamefont {F.}~\bibnamefont
  {Prati}}, \bibinfo {author} {\bibfnamefont {A.~R.}\ \bibnamefont {Renslo}},
  \bibinfo {author} {\bibfnamefont {H.~L.}\ \bibnamefont {Woodcock}}, \ and\
  \bibinfo {author} {\bibfnamefont {Y.}~\bibnamefont {Chen}},\ }\href@noop {}
  {\bibfield  {journal} {\bibinfo  {journal} {Journal of the American Chemical
  Society}\ }\textbf {\bibinfo {volume} {137}},\ \bibinfo {pages} {8086}
  (\bibinfo {year} {2015})}\BibitemShut {NoStop}%
\bibitem [{\citenamefont {Harris}\ \emph {et~al.}(2000)\citenamefont {Harris},
  \citenamefont {Zhao},\ and\ \citenamefont {Mildvan}}]{Harris00}%
  \BibitemOpen
  \bibfield  {author} {\bibinfo {author} {\bibfnamefont {T.}~\bibnamefont
  {Harris}}, \bibinfo {author} {\bibfnamefont {Q.}~\bibnamefont {Zhao}}, \ and\
  \bibinfo {author} {\bibfnamefont {A.}~\bibnamefont {Mildvan}},\ }\href@noop
  {} {\bibfield  {journal} {\bibinfo  {journal} {Journal of Molecular
  Structure}\ }\textbf {\bibinfo {volume} {552}},\ \bibinfo {pages} {97}
  (\bibinfo {year} {2000})}\BibitemShut {NoStop}%
\bibitem [{Note1()}]{Note1}%
  \BibitemOpen
  \bibinfo {note} {One might also consider whether the fact that X-ray crystal
  structures are refined with classical molecular dynamics using force fields
  that are parametrised for weak H-bonds may also be a problem. Such
  refinements may naturally bias towards weak bonds, i.e., the longer bond
  lengths that are common in proteins.}\BibitemShut {Stop}%
\bibitem [{\citenamefont {Kreevoy}\ and\ \citenamefont
  {Liang}(1980)}]{Kreevoy80}%
  \BibitemOpen
  \bibfield  {author} {\bibinfo {author} {\bibfnamefont {M.}~\bibnamefont
  {Kreevoy}}\ and\ \bibinfo {author} {\bibfnamefont {T.}~\bibnamefont
  {Liang}},\ }\href@noop {} {\bibfield  {journal} {\bibinfo  {journal} {Journal
  of the American Chemical Society}\ }\textbf {\bibinfo {volume} {102}},\
  \bibinfo {pages} {3315} (\bibinfo {year} {1980})}\BibitemShut {NoStop}%
\bibitem [{\citenamefont {McKenzie}\ \emph {et~al.}(2014)\citenamefont
  {McKenzie}, \citenamefont {Bekker}, \citenamefont {Athokpam},\ and\
  \citenamefont {Ramesh}}]{McKenzieJCP14}%
  \BibitemOpen
  \bibfield  {author} {\bibinfo {author} {\bibfnamefont {R.~H.}\ \bibnamefont
  {McKenzie}}, \bibinfo {author} {\bibfnamefont {C.}~\bibnamefont {Bekker}},
  \bibinfo {author} {\bibfnamefont {B.}~\bibnamefont {Athokpam}}, \ and\
  \bibinfo {author} {\bibfnamefont {S.~G.}\ \bibnamefont {Ramesh}},\ }\href
  {\doibase http://dx.doi.org/10.1063/1.4873352} {\bibfield  {journal}
  {\bibinfo  {journal} {The Journal of Chemical Physics}\ }\textbf {\bibinfo
  {volume} {140}},\ \bibinfo {pages} {174508} (\bibinfo {year}
  {2014})}\BibitemShut {NoStop}%
\bibitem [{\citenamefont {Hibbert}\ and\ \citenamefont
  {Emsley}(1990)}]{Hibbert90}%
  \BibitemOpen
  \bibfield  {author} {\bibinfo {author} {\bibfnamefont {F.}~\bibnamefont
  {Hibbert}}\ and\ \bibinfo {author} {\bibfnamefont {J.}~\bibnamefont
  {Emsley}},\ }\href@noop {} {\bibfield  {journal} {\bibinfo  {journal}
  {Advances in Physical Organic Chemistry}\ }\textbf {\bibinfo {volume} {26}},\
  \bibinfo {pages} {255} (\bibinfo {year} {1990})}\BibitemShut {NoStop}%
\bibitem [{\citenamefont {Mildvan}\ \emph {et~al.}(2002)\citenamefont
  {Mildvan}, \citenamefont {Massiah}, \citenamefont {Harris}, \citenamefont
  {Marks}, \citenamefont {Harrison}, \citenamefont {Viragh}, \citenamefont
  {Reddy},\ and\ \citenamefont {Kovach}}]{Mildvan02}%
  \BibitemOpen
  \bibfield  {author} {\bibinfo {author} {\bibfnamefont {A.~S.}\ \bibnamefont
  {Mildvan}}, \bibinfo {author} {\bibfnamefont {M.~A.}\ \bibnamefont
  {Massiah}}, \bibinfo {author} {\bibfnamefont {T.~K.}\ \bibnamefont {Harris}},
  \bibinfo {author} {\bibfnamefont {G.~T.}\ \bibnamefont {Marks}}, \bibinfo
  {author} {\bibfnamefont {D.~H.~T.}\ \bibnamefont {Harrison}}, \bibinfo
  {author} {\bibfnamefont {C.}~\bibnamefont {Viragh}}, \bibinfo {author}
  {\bibfnamefont {P.~M.}\ \bibnamefont {Reddy}}, \ and\ \bibinfo {author}
  {\bibfnamefont {I.~M.}\ \bibnamefont {Kovach}},\ }\href@noop {} {\bibfield
  {journal} {\bibinfo  {journal} {Journal of Molecular Structure}\ }\textbf
  {\bibinfo {volume} {615}},\ \bibinfo {pages} {163} (\bibinfo {year}
  {2002})}\BibitemShut {NoStop}%
\bibitem [{\citenamefont {McKenzie}(2012)}]{McKenzieCPL}%
  \BibitemOpen
  \bibfield  {author} {\bibinfo {author} {\bibfnamefont {R.~H.}\ \bibnamefont
  {McKenzie}},\ }\href {\doibase
  http://dx.doi.org/10.1016/j.cplett.2012.03.064} {\bibfield  {journal}
  {\bibinfo  {journal} {Chemical Physics Letters}\ }\textbf {\bibinfo {volume}
  {535}},\ \bibinfo {pages} {196 } (\bibinfo {year} {2012})}\BibitemShut
  {NoStop}%
\bibitem [{\citenamefont {Gilli}\ and\ \citenamefont {Gilli}(2009)}]{Gilli}%
  \BibitemOpen
  \bibfield  {author} {\bibinfo {author} {\bibfnamefont {G.}~\bibnamefont
  {Gilli}}\ and\ \bibinfo {author} {\bibfnamefont {P.}~\bibnamefont {Gilli}},\
  }\href@noop {} {\emph {\bibinfo {title} {The Nature of the Hydrogen Bond}}}\
  (\bibinfo  {publisher} {Oxford U.P., Oxford},\ \bibinfo {year}
  {2009})\BibitemShut {NoStop}%
\bibitem [{\citenamefont {Wang}\ \emph
  {et~al.}(2014{\natexlab{a}})\citenamefont {Wang}, \citenamefont {Fried},
  \citenamefont {Boxer},\ and\ \citenamefont {Markland}}]{Wang2014PNAS}%
  \BibitemOpen
  \bibfield  {author} {\bibinfo {author} {\bibfnamefont {L.}~\bibnamefont
  {Wang}}, \bibinfo {author} {\bibfnamefont {S.~D.}\ \bibnamefont {Fried}},
  \bibinfo {author} {\bibfnamefont {S.~G.}\ \bibnamefont {Boxer}}, \ and\
  \bibinfo {author} {\bibfnamefont {T.~E.}\ \bibnamefont {Markland}},\
  }\href@noop {} {\bibfield  {journal} {\bibinfo  {journal} {Proceedings of the
  National Academy of Sciences}\ }\textbf {\bibinfo {volume} {111}},\ \bibinfo
  {pages} {18454} (\bibinfo {year} {2014}{\natexlab{a}})}\BibitemShut {NoStop}%
\bibitem [{\citenamefont {Kanematsu}\ and\ \citenamefont
  {Tachikawa}(2014)}]{Kanematsu2014}%
  \BibitemOpen
  \bibfield  {author} {\bibinfo {author} {\bibfnamefont {Y.}~\bibnamefont
  {Kanematsu}}\ and\ \bibinfo {author} {\bibfnamefont {M.}~\bibnamefont
  {Tachikawa}},\ }\href@noop {} {\bibfield  {journal} {\bibinfo  {journal} {The
  Journal of Chemical Physics}\ }\textbf {\bibinfo {volume} {141}},\ \bibinfo
  {pages} {185101} (\bibinfo {year} {2014})}\BibitemShut {NoStop}%
\bibitem [{\citenamefont {Oltrogge}\ and\ \citenamefont
  {Boxer}(2015)}]{Oltrogge}%
  \BibitemOpen
  \bibfield  {author} {\bibinfo {author} {\bibfnamefont {L.~M.}\ \bibnamefont
  {Oltrogge}}\ and\ \bibinfo {author} {\bibfnamefont {S.~G.}\ \bibnamefont
  {Boxer}},\ }\href@noop {} {\bibfield  {journal} {\bibinfo  {journal} {ACS
  Central Science}\ }\textbf {\bibinfo {volume} {1}},\ \bibinfo {pages} {148}
  (\bibinfo {year} {2015})}\BibitemShut {NoStop}%
\bibitem [{\citenamefont {McKenzie}(2014)}]{McKenzieJCP}%
  \BibitemOpen
  \bibfield  {author} {\bibinfo {author} {\bibfnamefont {R.~H.}\ \bibnamefont
  {McKenzie}},\ }\href@noop {} {\bibfield  {journal} {\bibinfo  {journal} {The
  Journal of Chemical Physics}\ }\textbf {\bibinfo {volume} {141}},\ \bibinfo
  {pages} {104314} (\bibinfo {year} {2014})}\BibitemShut {NoStop}%
\bibitem [{\citenamefont {Bao}\ \emph {et~al.}(1999)\citenamefont {Bao},
  \citenamefont {Huskey}, \citenamefont {Kettner},\ and\ \citenamefont
  {Jordan}}]{Bao99}%
  \BibitemOpen
  \bibfield  {author} {\bibinfo {author} {\bibfnamefont {D.}~\bibnamefont
  {Bao}}, \bibinfo {author} {\bibfnamefont {W.~P.}\ \bibnamefont {Huskey}},
  \bibinfo {author} {\bibfnamefont {C.~A.}\ \bibnamefont {Kettner}}, \ and\
  \bibinfo {author} {\bibfnamefont {F.}~\bibnamefont {Jordan}},\ }\href@noop {}
  {\bibfield  {journal} {\bibinfo  {journal} {Journal of the American Chemical
  Society}\ }\textbf {\bibinfo {volume} {121}},\ \bibinfo {pages} {4684}
  (\bibinfo {year} {1999})}\BibitemShut {NoStop}%
\bibitem [{\citenamefont {Colbert}\ and\ \citenamefont
  {Miller}(1992)}]{Colbert92}%
  \BibitemOpen
  \bibfield  {author} {\bibinfo {author} {\bibfnamefont {D.~T.}\ \bibnamefont
  {Colbert}}\ and\ \bibinfo {author} {\bibfnamefont {W.~H.}\ \bibnamefont
  {Miller}},\ }\href@noop {} {\bibfield  {journal} {\bibinfo  {journal} {The
  Journal of Chemical Physics}\ }\textbf {\bibinfo {volume} {96}},\ \bibinfo
  {pages} {1982} (\bibinfo {year} {1992})}\BibitemShut {NoStop}%
\bibitem [{\citenamefont {Ichikawa}(2000)}]{Ichikawa00}%
  \BibitemOpen
  \bibfield  {author} {\bibinfo {author} {\bibfnamefont {M.}~\bibnamefont
  {Ichikawa}},\ }\href@noop {} {\bibfield  {journal} {\bibinfo  {journal}
  {Journal of Molecular Structure}\ }\textbf {\bibinfo {volume} {552}},\
  \bibinfo {pages} {63} (\bibinfo {year} {2000})}\BibitemShut {NoStop}%
\bibitem [{\citenamefont {Li}\ \emph {et~al.}(2011)\citenamefont {Li},
  \citenamefont {Walker},\ and\ \citenamefont {Michaelides}}]{Li11}%
  \BibitemOpen
  \bibfield  {author} {\bibinfo {author} {\bibfnamefont {X.-Z.}\ \bibnamefont
  {Li}}, \bibinfo {author} {\bibfnamefont {B.}~\bibnamefont {Walker}}, \ and\
  \bibinfo {author} {\bibfnamefont {A.}~\bibnamefont {Michaelides}},\
  }\href@noop {} {\bibfield  {journal} {\bibinfo  {journal} {Proceedings of the
  National Academy of Sciences}\ }\textbf {\bibinfo {volume} {108}},\ \bibinfo
  {pages} {6369} (\bibinfo {year} {2011})}\BibitemShut {NoStop}%
\bibitem [{\citenamefont {Habershon}\ \emph {et~al.}(2009)\citenamefont
  {Habershon}, \citenamefont {Markland},\ and\ \citenamefont
  {Manolopoulos}}]{Habershon09}%
  \BibitemOpen
  \bibfield  {author} {\bibinfo {author} {\bibfnamefont {S.}~\bibnamefont
  {Habershon}}, \bibinfo {author} {\bibfnamefont {T.~E.}\ \bibnamefont
  {Markland}}, \ and\ \bibinfo {author} {\bibfnamefont {D.~E.}\ \bibnamefont
  {Manolopoulos}},\ }\href@noop {} {\bibfield  {journal} {\bibinfo  {journal}
  {The Journal of Chemical Physics}\ }\textbf {\bibinfo {volume} {131}},\
  \bibinfo {pages} {024501} (\bibinfo {year} {2009})}\BibitemShut {NoStop}%
\bibitem [{\citenamefont {Markland}\ and\ \citenamefont
  {Berne}(2012)}]{Markland12}%
  \BibitemOpen
  \bibfield  {author} {\bibinfo {author} {\bibfnamefont {T.}~\bibnamefont
  {Markland}}\ and\ \bibinfo {author} {\bibfnamefont {B.}~\bibnamefont
  {Berne}},\ }\href@noop {} {\bibfield  {journal} {\bibinfo  {journal}
  {Proceedings of the National Academy of Sciences}\ }\textbf {\bibinfo
  {volume} {109}},\ \bibinfo {pages} {7988} (\bibinfo {year}
  {2012})}\BibitemShut {NoStop}%
\bibitem [{\citenamefont {Romanelli}\ \emph {et~al.}(2013)\citenamefont
  {Romanelli}, \citenamefont {Ceriotti}, \citenamefont {Manolopoulos},
  \citenamefont {Pantalei}, \citenamefont {Senesi},\ and\ \citenamefont
  {Andreani}}]{Romanelli13}%
  \BibitemOpen
  \bibfield  {author} {\bibinfo {author} {\bibfnamefont {G.}~\bibnamefont
  {Romanelli}}, \bibinfo {author} {\bibfnamefont {M.}~\bibnamefont {Ceriotti}},
  \bibinfo {author} {\bibfnamefont {D.~E.}\ \bibnamefont {Manolopoulos}},
  \bibinfo {author} {\bibfnamefont {C.}~\bibnamefont {Pantalei}}, \bibinfo
  {author} {\bibfnamefont {R.}~\bibnamefont {Senesi}}, \ and\ \bibinfo {author}
  {\bibfnamefont {C.}~\bibnamefont {Andreani}},\ }\href@noop {} {\bibfield
  {journal} {\bibinfo  {journal} {The Journal of Physical Chemistry Letters}\
  }\textbf {\bibinfo {volume} {4}},\ \bibinfo {pages} {3251} (\bibinfo {year}
  {2013})}\BibitemShut {NoStop}%
\bibitem [{\citenamefont {Wang}\ \emph
  {et~al.}(2014{\natexlab{b}})\citenamefont {Wang}, \citenamefont {Ceriotti},\
  and\ \citenamefont {Markland}}]{WangJCP14}%
  \BibitemOpen
  \bibfield  {author} {\bibinfo {author} {\bibfnamefont {L.}~\bibnamefont
  {Wang}}, \bibinfo {author} {\bibfnamefont {M.}~\bibnamefont {Ceriotti}}, \
  and\ \bibinfo {author} {\bibfnamefont {T.~E.}\ \bibnamefont {Markland}},\
  }\href@noop {} {\bibfield  {journal} {\bibinfo  {journal} {The Journal of
  Chemical Physics}\ }\textbf {\bibinfo {volume} {141}},\ \bibinfo {pages}
  {104502} (\bibinfo {year} {2014}{\natexlab{b}})}\BibitemShut {NoStop}%
\bibitem [{\citenamefont {Edison}\ \emph {et~al.}(1995)\citenamefont {Edison},
  \citenamefont {Weinhold},\ and\ \citenamefont {Markley}}]{Edison95}%
  \BibitemOpen
  \bibfield  {author} {\bibinfo {author} {\bibfnamefont {A.~S.}\ \bibnamefont
  {Edison}}, \bibinfo {author} {\bibfnamefont {F.}~\bibnamefont {Weinhold}}, \
  and\ \bibinfo {author} {\bibfnamefont {J.~L.}\ \bibnamefont {Markley}},\
  }\href@noop {} {\bibfield  {journal} {\bibinfo  {journal} {Journal of the
  American Chemical Society}\ }\textbf {\bibinfo {volume} {117}},\ \bibinfo
  {pages} {9619} (\bibinfo {year} {1995})}\BibitemShut {NoStop}%
\bibitem [{\citenamefont {Bothma}\ \emph {et~al.}(2010)\citenamefont {Bothma},
  \citenamefont {Gilmore},\ and\ \citenamefont {McKenzie}}]{Bothma10}%
  \BibitemOpen
  \bibfield  {author} {\bibinfo {author} {\bibfnamefont {J.~P.}\ \bibnamefont
  {Bothma}}, \bibinfo {author} {\bibfnamefont {J.~B.}\ \bibnamefont {Gilmore}},
  \ and\ \bibinfo {author} {\bibfnamefont {R.~H.}\ \bibnamefont {McKenzie}},\
  }\href@noop {} {\bibfield  {journal} {\bibinfo  {journal} {New Journal of
  Physics}\ }\textbf {\bibinfo {volume} {12}},\ \bibinfo {pages} {055002}
  (\bibinfo {year} {2010})}\BibitemShut {NoStop}%
\bibitem [{\citenamefont {Klug}\ \emph {et~al.}(1997)\citenamefont {Klug},
  \citenamefont {Lee}, \citenamefont {Lee}, \citenamefont {Kreevoy},
  \citenamefont {Yaris},\ and\ \citenamefont {Schaefer}}]{Klug97}%
  \BibitemOpen
  \bibfield  {author} {\bibinfo {author} {\bibfnamefont {C.~A.}\ \bibnamefont
  {Klug}}, \bibinfo {author} {\bibfnamefont {P.~L.}\ \bibnamefont {Lee}},
  \bibinfo {author} {\bibfnamefont {I.~S.~H.}\ \bibnamefont {Lee}}, \bibinfo
  {author} {\bibfnamefont {M.~M.}\ \bibnamefont {Kreevoy}}, \bibinfo {author}
  {\bibfnamefont {R.}~\bibnamefont {Yaris}}, \ and\ \bibinfo {author}
  {\bibfnamefont {J.}~\bibnamefont {Schaefer}},\ }\href@noop {} {\bibfield
  {journal} {\bibinfo  {journal} {The Journal of Physical Chemistry B}\
  }\textbf {\bibinfo {volume} {101}},\ \bibinfo {pages} {8086} (\bibinfo {year}
  {1997})}\BibitemShut {NoStop}%
\bibitem [{\citenamefont {Loh}\ and\ \citenamefont {Markley}(1994)}]{Loh94}%
  \BibitemOpen
  \bibfield  {author} {\bibinfo {author} {\bibfnamefont {S.~N.}\ \bibnamefont
  {Loh}}\ and\ \bibinfo {author} {\bibfnamefont {J.~L.}\ \bibnamefont
  {Markley}},\ }\href@noop {} {\bibfield  {journal} {\bibinfo  {journal}
  {Biochemistry}\ }\textbf {\bibinfo {volume} {33}},\ \bibinfo {pages} {1029}
  (\bibinfo {year} {1994})}\BibitemShut {NoStop}%
\bibitem [{\citenamefont {Khare}\ \emph {et~al.}(1999)\citenamefont {Khare},
  \citenamefont {Alexander},\ and\ \citenamefont {Orban}}]{Khare1999}%
  \BibitemOpen
  \bibfield  {author} {\bibinfo {author} {\bibfnamefont {D.}~\bibnamefont
  {Khare}}, \bibinfo {author} {\bibfnamefont {P.}~\bibnamefont {Alexander}}, \
  and\ \bibinfo {author} {\bibfnamefont {J.}~\bibnamefont {Orban}},\
  }\href@noop {} {\bibfield  {journal} {\bibinfo  {journal} {Biochemistry}\
  }\textbf {\bibinfo {volume} {38}},\ \bibinfo {pages} {3918} (\bibinfo {year}
  {1999})}\BibitemShut {NoStop}%
\bibitem [{\citenamefont {LiWang}\ and\ \citenamefont
  {Bax}(1996)}]{Liwang1996}%
  \BibitemOpen
  \bibfield  {author} {\bibinfo {author} {\bibfnamefont {A.~C.}\ \bibnamefont
  {LiWang}}\ and\ \bibinfo {author} {\bibfnamefont {A.}~\bibnamefont {Bax}},\
  }\href@noop {} {\bibfield  {journal} {\bibinfo  {journal} {Journal of the
  American Chemical Society}\ }\textbf {\bibinfo {volume} {118}},\ \bibinfo
  {pages} {12864} (\bibinfo {year} {1996})}\BibitemShut {NoStop}%
\bibitem [{\citenamefont {Li}\ \emph {et~al.}(2015)\citenamefont {Li},
  \citenamefont {Srivastava}, \citenamefont {Kim}, \citenamefont {Taylor},\
  and\ \citenamefont {Veglia}}]{Li2015}%
  \BibitemOpen
  \bibfield  {author} {\bibinfo {author} {\bibfnamefont {G.~C.}\ \bibnamefont
  {Li}}, \bibinfo {author} {\bibfnamefont {A.~K.}\ \bibnamefont {Srivastava}},
  \bibinfo {author} {\bibfnamefont {J.}~\bibnamefont {Kim}}, \bibinfo {author}
  {\bibfnamefont {S.~S.}\ \bibnamefont {Taylor}}, \ and\ \bibinfo {author}
  {\bibfnamefont {G.}~\bibnamefont {Veglia}},\ }\href@noop {} {\bibfield
  {journal} {\bibinfo  {journal} {Biochemistry}\ }\textbf {\bibinfo {volume}
  {54}},\ \bibinfo {pages} {4042} (\bibinfo {year} {2015})}\BibitemShut
  {NoStop}%
\bibitem [{\citenamefont {Thakur}\ \emph {et~al.}(2013)\citenamefont {Thakur},
  \citenamefont {Chandra}, \citenamefont {Dubey}, \citenamefont {D'Silva},\
  and\ \citenamefont {Atreya}}]{Thakur13}%
  \BibitemOpen
  \bibfield  {author} {\bibinfo {author} {\bibfnamefont {A.}~\bibnamefont
  {Thakur}}, \bibinfo {author} {\bibfnamefont {K.}~\bibnamefont {Chandra}},
  \bibinfo {author} {\bibfnamefont {A.}~\bibnamefont {Dubey}}, \bibinfo
  {author} {\bibfnamefont {P.}~\bibnamefont {D'Silva}}, \ and\ \bibinfo
  {author} {\bibfnamefont {H.~S.}\ \bibnamefont {Atreya}},\ }\href@noop {}
  {\bibfield  {journal} {\bibinfo  {journal} {Angewandte Chemie International
  Edition}\ }\textbf {\bibinfo {volume} {52}},\ \bibinfo {pages} {2440}
  (\bibinfo {year} {2013})}\BibitemShut {NoStop}%
\end{thebibliography}%

\end{document}